\RequirePackage{ifpdf}

\documentclass[aps,onecolumn,notitlepage,amsmath,amssymb,preprintnumbers,nofootinbib,tightenlines,nobibnotes,superscriptaddress]{revtex4-1}

\ifpdf
  \usepackage{color}
\fi

\usepackage{graphicx}
\usepackage{amsmath}
\usepackage{amsfonts}
\usepackage{amssymb}
\usepackage{hyperref}
\usepackage{xcolor}

\usepackage{IEEEtrantools}

\newcommand{\bC}{\mathbb{C}}

\definecolor{MyGrey}{rgb}{0.5,0.5,0.5}

\begin{document}


\title{On rigid supersymmetry and notions of holomorphy in five dimensions}

\author{Yiwen Pan}
\email{yiwen.pan@stonybrook.edu}
\affiliation{C. N. Yang Institute for Theoretical Physics, Stony Brook University, Stony Brook, NY 11794}

\author{Johannes Schmude}
\email{schmudejohannes@uniovi.es}
\affiliation{Department of Physics, Universidad de Oviedo, 33007, Oviedo, Spain}

\begin{abstract}
We study the equations governing rigid $\mathcal{N}=1$ supersymmetry in five dimensions. If the supersymmetry spinor satisfies a reality condition, these are foliations admitting families of almost complex structures on the leaves. In other words, all these manifolds have families of almost Cauchy-Riemann (CR) structures. After deriving integrability conditions under which circumstances the almost CR structure defines a CR manifold or a transversally holomorphic foliation (THF), we discuss implications on localization. We also discuss potential global obstructions to the existence of solutions.
\end{abstract}

\maketitle

\section{Introduction}
\label{sec:introduction}

The work of Festuccia and Seiberg \cite{Festuccia:2011ws} provides an easy to implement recipe to construct gauge theories with rigid supersymmetry on curved manifolds. One picks an (off-shell) supergravity theory coupled to the vector and matter multiplets and solves the Killing spinor equations arising from the gravitino and dilatino variations. These in turn impose constraints on the auxiliary fields that then appear as parameters in the lagrangian of the field theory once one has decoupled gravity using a suitable scaling limit. Using supergravity techniques, the approach allows for a classification of admissible spacetime manifolds \cite{Dumitrescu:2012ha,Closset:2012ru,Closset:2013vra,Closset:2014uda,Cassani:2012ri,Klare:2012gn,Klare:2013dka}.

In five dimensions, initial progress came not so much from the rigid limit of supergravity, as from the direct construction of supersymmetric field theories on increasingly intricate manifolds \cite{Hosomichi:2012ek,Kallen:2012cs,Qiu:2013pta,Qiu:2014cha}. Here, a key observation was that the theory on the five-sphere can be readily generalized to generic Sasaki-Einstein manifolds and even to manifolds that support only a K-contact structure. Contact with the formalism of the rigid limit was made by Imamura and Matsuno \cite{Imamura:2014ima} who solved the Killing spinor equations of the $\mathcal{N}=1$ supergravity of Kugo, Ohashi and Zucker \cite{Kugo:2000hn,Kugo:2000af,Zucker:2000uq} locally, as well as in \cite{Pan:2013uoa} where the equations arising from the gravitino were discussed in the context of the same supergravity theory. For maximal supergravity in $d=5$, consider \cite{Cordova:2013kx}; a superspace approach to five-dimensional $\mathcal{N}=1$ backgrounds has been developed in \cite{Kuzenko:2008wr,Kuzenko:2014eqa}.

Of course, the interest in rigid supersymmetry goes hand in hand with the idea of localization following \cite{Pestun:2007rz}. In this context, the Sasakian case has proven to be interesting. Sasakian manifolds, being the odd dimensional cousin of K\"ahler ones, support something akin to an integrable complex structure on the space transverse to the Reeb vector $R$. That is, the tangent bundle admits the decomposition
\begin{equation}\label{eq:almost-CR-structure}
   T_{\bC}M = T^{1,0} \oplus T^{0,1} \oplus \bC R.
\end{equation}
This is integrable in a sense that we will discuss shortly and it follows that one can introduce differentials $\partial_b$ and $\bar{\partial}_b$ that correspond to the Dolbeault operators $\partial$ and $\bar{\partial}$ that are familiar from complex geometry. In \cite{Schmude:2014lfa,Pan:2014bwa} this was used in order to simplify the calculation of the perturbative part of the supersymmetric partition function and to solve the BPS equations on the Higgs branch.

This note focusses on the question to what extend this notion and use of holomorphy can be extended to general five-dimensional backgrounds admitting rigid $\mathcal{N}=1$ supersymmetry. Our analysis is based on the gravitino and dilatino equations of \cite{Kugo:2000hn,Kugo:2000af} which in our conventions and in Euclidean signature are
  \begin{equation}
    {D _m}{\xi _I} = {t_I}^J{\Gamma _m}{\xi _J} + {\mathcal{F}_{mn}}{\Gamma ^n}{\xi _I} + \frac{1}{2}{\mathcal{V}^{pq}}{\Gamma _{mpq}}{\xi _I}
    \label{Killing}
  \end{equation}
and
  \begin{equation}
      4\left[ {\left( {{D _m}{t_I}^J} \right){\Gamma ^m} + {t_I}^J{{\left( {\mathcal{F} + 2 \mathcal{V}} \right)}_{mn}}{\Gamma ^{mn}}} \right]{\xi _J} + \left( {4{\nabla _m}{\mathcal{V}^{mn}}{\Gamma _n} + {\mathcal{F}_{mn}}{\mathcal{F}_{kl}}{\Gamma ^{mnkl}} + C} \right){\xi _I} = 0
      \label{eq:Dilatino}
  \end{equation}
Here, $I = 1, 2$ are indices for the fundamental representation of $SU(2)_\mathcal{R}$. $\mathcal{F} = d \mathcal{A}$ is a $U(1)$ field strength and $\mathcal{V}$ an antisymmetric tensor. The triplet $t_I^{\phantom{I}J}$ is valued in the adjoint representation of $SU(2)_\mathcal{R}$. The covariant derivatives are $D_m \xi_I = \nabla_m \xi_I - A_{mI}^{\phantom{mI}J} \xi_J$ and $D_m t_I^{\phantom{I}J} = \nabla_m t_I^{\phantom{I}J} - [A_m, t]_I^{\phantom{I}J}$. For later convenience, note that \eqref{Killing} can also be rewritten as
  \begin{equation}
    {D_m}{\xi _I} = {\Gamma _m}{{\tilde \xi }_I} + \frac{1}{2}\left( {{\mathcal{V}^{pq}} - {\mathcal{F}^{pq}}} \right){\Gamma _{mpq}}{\xi _I}, \;\;\;\;{{\tilde \xi }_I} = {t_I}^J{\xi _J} + \frac{1}{2}{\mathcal{F}_{mn}}{\Gamma ^{mn}}{\xi _I}.
  \end{equation}

In the Lorentzian theory, the spinors $\xi_I$ satisfy a symplectic Majorana condition \eqref{eq:symplectic_Majorana_condition}. Transitioning to the Euclidean theory one usually drops such reality conditions and effectively doubles the degrees of freedom of all fields involved. In general, the spinor $\xi_I$ defines a possibly complex vector $R$. Imposing the reality condition \eqref{eq:symplectic_Majorana_condition} for $\xi_I$ it follows that $R$ is real and non-vanishing and that the tangent space decomposes as in \eqref{eq:almost-CR-structure}, which one refers to as an \emph{almost Cauchy-Riemann (CR) structure (of hypersurface type)} \cite{dragomir2007differential}. 

In opposite to the familiar case of almost complex structures, the integrability condition for \eqref{eq:almost-CR-structure} is not unique. Indeed, there are two possibilities. To begin, there is the case of a \emph{integrable CR structure},
\begin{equation}\label{eq:CR_integrability_condition}
  [T^{1,0}, T^{1,0}] \subseteq T^{1,0},
\end{equation}
that defines a \emph{CR manifold}. CR manifolds have previously appeared in the context of the rigid limit of new minimal supergravity with Lorentzian signature in \cite{Cassani:2012ri}; there, the authors found fibrations of the real line over three dimensional CR manifolds. Alternatively, there is the condition
\begin{equation}\label{eq:THF_integrability_condition}
  [T^{1,0} \oplus \bC R,T^{1,0} \oplus \bC R] \subseteq T^{1,0} \oplus \bC R,
\end{equation}
which defines a \emph{transversally holomorphic foliation (THF)}.\footnote{For some background material on transversely holomorphic foliations, see e.g.~\cite{biswas2001transversely,jacobowit2000transversel}.} The work of \cite{Dumitrescu:2012ha} relates rigid supersymmetry in three dimensions with the existence of a THF. Note that Sasakian manifolds fulfill both \eqref{eq:CR_integrability_condition} and \eqref{eq:THF_integrability_condition} as here $[R_{\text{Sasakian}}, T^{1,0}] \subseteq T^{1,0}$.

Naturally, the question whether solutions to the Killing spinor equations \eqref{Killing} and \eqref{eq:Dilatino} admit integrable CR structures or THFs is closely related to the question whether a given five dimensional manifold $\mathcal{M}$ admits any solution in the first place. As we alluded above, this question was already addressed in \cite{Imamura:2014ima} and \cite{Pan:2013uoa}, but not exhaustively answered. As we will see, existence of a solution to the Killing spinor equations that satisfies the symplectic Majorana condition implies the existence of a globally non-vanishing Killing vector field parallel to $R$. We will show that the existence of such a Killing vector field is not only necessary, but also sufficient. While we will do so by directly constructing a single solution and arguing that there are no topological obstructions, one can already give a short argument why one should be able to expect this result. The existence of a non-vanishing vector field implies that $\mathcal{M}$ admits an $SO(4)$ structure. Since the theory has an $SU(2)_\mathcal{R}$ symmetry, one can perform an operation akin to a Witten twist in four dimensions and identify the the $SU(2)_{\mathcal{R}}$ with an $SU(2)$ factor inside the structure group.

The structure of this note is as follows: The relation between the supersymmetry spinor $\xi_I$, almost CR-structures and almost contact structures is the topic of section \ref{sec:algebraic_properties}. Then, we will discuss the integrability of the Killing spinor equations, possible obstructions and general differential properties of \eqref{Killing} and \eqref{eq:Dilatino} in section \ref{sec:Differential_properties}. Section \ref{sec:implications_for_localization} is concerned with the implications for localization. We will argue that the results of \cite{Schmude:2014lfa,Pan:2013uoa} can be generalized to CR-manifolds and THFs. Subsequently we discuss the existence of globally well-defined solutions (section \ref{sec:krw_twist}) before concluding with some examples from the literature in section \ref{sec:examples}. Various appendices complement the discussion.

During the final stages of this project \cite{Alday:2015lta} appeared, which has some overlap with our work. There, the authors study rigid supersymmetry on Riemannian five-manifolds using a holographic approach.

\section{Algebraic Properties}
\label{sec:algebraic_properties}

In this section we will discuss the algebraic structures arrising from the existence of the spinors $\xi_I$.

\subsection{The Almost Contact Structure}

The bulk of our analysis is based on the following set of bi-spinors that can be defined for any given $\xi_I$:
  \begin{equation}\label{eq:bispinors}
    s \equiv {\epsilon ^{IJ}}\left( {{\xi _I}{\xi _J}} \right),\qquad
    {R^m} \equiv  - {s^{ - 1}}{\epsilon ^{IJ}}\left( {{\xi _I}{\Gamma ^m}{\xi _J}} \right) \equiv {g^{mn}}{\kappa _n},\qquad
    {\left( {{\Theta _{IJ}}} \right)_{mn}} \equiv \left( {{\xi _I}{\Gamma _{mn}}{\xi _J}} \right).
  \end{equation}
Let us emphasize the presence of the minus sign as well as the normalizing factor $s^{-1}$ in the definition of $R$ where we tacitly assume that $s \neq 0$. If one imposes the symplectic Majorana condition \eqref{eq:symplectic_Majorana_condition} one finds that $s$ and $R$ are a real function and a real vector field respectively. A further consequence of \eqref{eq:symplectic_Majorana_condition} is that $s \ge 0$ with equality if and only if $\xi_I = 0$. It follows that $s > 0$ everywhere on $M$ since the gravitino equation is linear and of first order. Finally, the two forms $\Theta_{IJ}$ lie in the adjoint representation of $SU(2)_\mathcal{R}$.

Using Fierz-identities, one can show the following identities involving the bispinors:
\begin{equation}\label{eq:bi-spinors_algebraic_properties}
  1 = \iota_R \kappa, \qquad
  0 = \iota_R \Theta_{IJ}, \qquad
  \iota _R * \Theta _{IJ} = \Theta _{IJ}, \qquad
  \star \Theta_{IJ} = \kappa \wedge \Theta_{IJ}, \qquad
  R^m \gamma_m \xi_I = - \xi_I.
\end{equation}
Here, $*$ is the usual five-dimensional Hodge dual and $\iota_R$ denotes interior multiplication. The first of the above equations tells us that $M$ carries an $SO(4)$ structure. This allows us to introduce a lot of structure that is familiar from four-dimensional geometry. As is usual, we will refer to vectors and forms parallel to $R$ and $\kappa$ respectively as vertical and their orthogonal complement as horizontal. I.e.~forms can be decomposed as $\omega = \omega_H + \omega_V$. Then the Hodge dual defines the notion of self-dual and anti self-dual forms on the horizontal subspace. See equations \eqref{eq:horizontal-vertical_decomposition} and \eqref{eq:self_duality_decomposition} for full definitions. Since the $\Theta_{IJ}$ are both horizontal and self-dual, $\Theta_{IJ} = (\Theta_{IJ})^+$, they define an isomorphism between $su(2)_\mathcal{R}$ and the $su(2)_+$ factor in the typical $so(4) \cong su(2)_+ \times su(2)_-$ decomposition of the Lie algebra of the structure group. One can also verify some more involved identities involving $\Theta_{IJ}$:
\begin{equation}\label{eq:involved_theta_identities}
  \begin{aligned}
    \Theta_{IJmp} \Theta_{KL}^{\phantom{KL}pn} &= - \frac{1}{4} s^2 (\epsilon_{IK} \epsilon_{JL} + \epsilon_{IL} \epsilon_{JK}) \Pi_m^{\phantom{m}n} + \frac{1}{4} s ( \epsilon_{JK} \Theta_{ILm}^{\phantom{ILm}n} + \epsilon_{IK} \Theta_{JLm}^{\phantom{JLm}n} + \epsilon_{JL} \Theta_{IKm}^{\phantom{IKm}n} + \epsilon_{IL} \Theta_{JKm}^{\phantom{JKm}n}), \\
    s^{-2} \Theta_{IJkl} \Theta_{mn}^{IJ} &= \frac{1}{2} \left( \Pi_{km} \Pi_{ln} - \Pi_{kn} \Pi_{lm} + \epsilon_{klmnp} R^p \right).
  \end{aligned}
\end{equation}
Here $\Pi_{mn} = g_{mn} - \kappa_m \kappa_n$ and thus the latter of these is a projection to horizontal, self-dual two-forms.

Suppose now that $\mathfrak{m}_{IJ}$ is an $SU(2)_\mathcal{R}$ triplet. Later we will show that $\mathfrak{m}_{IJ} = t_{IJ}$ emerges naturally when imposing integrability and we will refer to this as the canonical choice. Yet for now, we continue with a generic $\mathfrak{m}_{IJ}$ and define\footnote{\label{fn:su(2)_determinant_identity}
Note that
\begin{equation*}
  \sum\limits_{IJ} {{\mathfrak{m}_I}^J} {\mathfrak{m}_J}^I = {\mathfrak{m}_1}^1{\mathfrak{m}_1}^1 + {\mathfrak{m}_1}^2{\mathfrak{m}_2}^1 + {\mathfrak{m}_2}^1{\mathfrak{m}_1}^2 + {\mathfrak{m}_2}^2{\mathfrak{m}_2}^2 =  - 2{\mathfrak{m}_{11}}{\mathfrak{m}_{22}} + 2{\mathfrak{m}_{12}}{\mathfrak{m}_{21}} =  - 2\det {\mathfrak{m}_{ \bullet  \bullet }}.
\end{equation*}
}
$ \det {\mathfrak{m}} \equiv -1/2 \sum\limits_{IJ} {{\mathfrak{m}_I}^J} {\mathfrak{m}_J}^I$. Once we impose the reality condition \eqref{eq:su2_triplet_mIJ_reality_condition} for $\mathfrak{m}_{IJ}$, $\det \mathfrak{m}$ will be positive semi-definite. For now we proceed with the milder assumption $\det \mathfrak{m} \ne 0$ and define the following tensor
  \begin{equation}
    {\Phi _{mn}} = (\Phi[\mathfrak{m}])_{mn} \equiv {s^{ - 1}}\sqrt {\frac{1}{{\det \mathfrak{m}}}} {\mathfrak{m}^{IJ}}{\left( {{\Theta _{IJ}}} \right)_{mn}}.
    \label{phi}
  \end{equation}
As follows from \eqref{eq:involved_theta_identities}, $\Phi$ satisfies the following condition:
  \begin{equation}\label{eq:phi-squared}
    {\Phi ^m}_k{\Phi ^k}_n =  - \delta _n^m + {R^m}{\kappa _n}.
  \end{equation}
Mathematicians refer to a multiplet $(\kappa ,R,\Phi )$ as an \emph{almost contact structure} if
  \begin{equation}\label{eq:almost_contact_structure}
    {\kappa _m}{R^m} = 1,\;\;\;\;{\Phi ^m}_k{\Phi ^k}_n =  - \delta _n^m + {R^m}{\kappa _n},\;\;\;\;{\Phi ^m}_n{R^n} = {\kappa _n}{\Phi ^n}_m = 0
  \end{equation}
As we have shown, the quantities defined using $\xi_I$ and a suitable $\mathfrak{m}_{IJ}$ satisfy these relations, and therefore define an almost contact structure. Note that $\Phi$ is invariant under $\mathfrak{m}_{IJ} \mapsto f \mathfrak{m}_{IJ}$ for any non-zero function $f$.

\subsection{The Almost CR Structure}

Equations \eqref{eq:phi-squared} and \eqref{eq:almost_contact_structure} indicate that for each $\mathfrak{m}$, $\Phi[\mathfrak{m}]$ defines an almost CR structure. Indeed, each $\Phi[\mathfrak{m}]$ induces a decomposition of the complexified tangent bundle as in \eqref{eq:almost-CR-structure} via
\begin{equation}
  X \in T^{1,0} \quad \Leftrightarrow \quad \Phi X = \imath X.
\end{equation}
The decomposition holds also for the exterior algebra and all horizontal $n$-forms $\omega = \omega_H$ can be decomposed into $(p,q)$-forms via
  \begin{equation}
    \omega  = \sum\limits_{p + q = n} {{\omega ^{p,q}}} .
  \end{equation}
In this context $\Phi_{mn}$ is a horizontal $(1,1)$-form. Similar to the case of four-dimensional K\"ahler manifolds, self-dual and anti-self-dual 2-forms have a simple $(p,q)$-decomposition,
  \begin{equation}
    {\omega ^ + } = {\omega ^{2,0}} + {\omega ^{0,2}} + {\left. \omega  \right|_\Phi },\qquad
    {\omega ^ - } = {\omega ^{1,1}},
  \end{equation}
with $\omega^{1,1}$ primitive and thus annihilated by contraction with $\Phi$.

We continue by discussing the integrability of the almost CR structure \eqref{eq:almost-CR-structure}. While this can be done using a direct analysis of the Niejenhuis tensor, we prefer to do a spinorial analysis in the spirit of \cite{Festuccia:2011ws}.\footnote{For a third possibility using differential forms orthogonal to $T^{1,0}$ or $T^{1,0} \oplus \mathbb{R}$ respectively see \cite{Cassani:2012ri}.}
This is computationally more straight forward, yet requires us to impose the reality condition
  \begin{equation}\label{eq:su2_triplet_mIJ_reality_condition}
    \overline {{\mathfrak{m}_{IJ}}}  = {\epsilon ^{II'}}{\epsilon ^{JJ'}}{\mathfrak{m}_{I'J'}}
  \end{equation}
for the triplet which we alluded to previously. The bar denotes complex conjugation. Let us emphasize that we are also using the symplectic Majorana condition since we assume $R$ to be real.

In appendix \ref{sec:appendix_spinorial_holomorphy_conditions} we show that one can characterize elements of $T^{1,0}$ in terms of a spinorial equation:
  \begin{equation}\label{spinorial-(1,0)}
    X \in T^{1,0} \quad \Leftrightarrow \quad {X^m}{H_I}^J{\Gamma _m}{\xi _J} = 0,
  \end{equation}
where
  \begin{equation}
    {H_I}^J = H_I^{\phantom{I}J}[\mathfrak{m}] = \sqrt {\frac{1}{{\det \mathfrak{m}}}} {\mathfrak{m}_I}^J - i\delta _I^J.
  \end{equation}
Similarly, one can also characterize the tangent vectors in $T^{1,0} \oplus \bC R$ by the spinorial equation
  \begin{equation}
    X \in T^{1,0} \oplus \bC R \quad \Leftrightarrow \quad \left(\Pi^m_{\phantom{m}n} {X^n}\right) {H_I}^J{\Gamma _m}{\xi _J} = 0.
    \label{spinorial-THF}
  \end{equation}
Recall that $\Pi^m_{\phantom{m}n} = \delta _n^m - {R^m}{\kappa _n}$ is a projection that maps a generic tangent vector to its horizontal component.

\section{Differential Properties}
\label{sec:Differential_properties}

We finally turn to the integrability conditions for the decomposition \eqref{eq:almost-CR-structure}. To do so, we will first establish some useful identities involving the bispinors \eqref{eq:bispinors} and the gravitino \eqref{Killing} and dilatino \eqref{eq:Dilatino} variations. Subsequently we consider the case of CR structures as a warm-up before studying the integrability conditions for THFs.

\subsection{Supersymmetry variations and bispinors}
\label{sec:susy_variations_and_bispinors}

Studying the gravitino variation \eqref{Killing}, one finds that the scalar $s$ satisfies ${\nabla _n}s = 2s{R^m}{\mathcal{F}_{mn}}$, from which it follows that $\mathcal{L}_R \mathcal{F} = \mathcal{L}_R s = 0$. Similarly, the non-normalized vector field $sR$ is Killing:
\begin{equation}
  {\nabla _m}\left( {s{R_n}} \right) = 2{\left( {{t^{IJ}}{\Theta _{IJ}}} \right)_{mn}} - 2s{\mathcal{F}_{mn}} - 2s{\left( {{\iota _R} \star \mathcal{V}} \right)_{mn}}, \;\;\;\;{\nabla _m}\left( {s{R_n}} \right) + {\nabla _n}\left( {s{R_m}} \right) = 0.
\end{equation}
One also finds that $\iota_R d\kappa = - s^{-1} ds$ while $\iota_R d(s\kappa) = - 2ds$.
There is a more involved relation involving the two-form $t^{IJ} \Theta_{IJ}$:
\begin{IEEEeqnarray}{rCl}\label{eq:Killing-3}
  {\nabla _k}{\left( {{t^{IJ}}{\Theta _{IJ}}} \right)_{mn}} &=& {D_k}{t^{IJ}}{\left( {{\Theta _{IJ}}} \right)_{mn}} + 2 \det t\left( {{g_{nk}}{R_m} - {g_{mk}}{R_n}} \right) + 2{\mathcal{F}_{kp}}{t^{IJ}}\left( {{\xi _I}{\Gamma _{mn}}^p{\xi _J}} \right) \nonumber \\
  &&+ {g_{nk}}{\mathcal{V}^{pq}}{t^{IJ}}\left( {{\xi _I}{\Gamma _{mpq}}{\xi _J}} \right) - 2{\mathcal{V}_n}^q{t^{IJ}}\left( {{\xi _I}{\Gamma _{mkq}}{\xi _J}} \right) \nonumber \\
  &&- {g_{mk}}{\mathcal{V}^{pq}}{t^{IJ}}\left( {{\xi _I}{\Gamma _{npq}}{\xi _J}} \right) + 2{\mathcal{V}_m}^q{t^{IJ}}\left( {{\xi _I}{\Gamma _{nkq}}{\xi _J}} \right). 
\end{IEEEeqnarray}

Similarly we are interested in the consequences of the dilatino equation \eqref{eq:Dilatino} for bispinors and background fields. By contraction with $t^{IJ}\xi_I$ one finds that $R^m \nabla_m ({t_{IJ}}{t^{IJ}} ) = 2 \mathcal{L}_R \det t = 0$. Contraction with $\xi^I$ on the other hand fixes the value of the scalar,
\begin{equation}\label{eq:scalar_C}
  C = 4{\kappa_n}{\nabla _m}{\mathcal{V}^{mn}} - 4{s^{ - 1}}{\left( {\mathcal{F} + 2\mathcal{V}} \right)_{mn}}{\left( {{t^{IJ}}{\Theta _{IJ}}} \right)^{mn}} + 2{\left( {{\iota _R}*\mathcal{F}} \right)^{mn}}{\mathcal{F}_{mn}}.
\end{equation}
We can extract additional information from the dilatino equation and start by projecting it onto its ``chiral'' components. Recalling the last identity in \eqref{eq:bi-spinors_algebraic_properties} we consider the projector $\frac{1}{2} (1 - R^m \Gamma_m)$. Acting on \eqref{eq:Dilatino} and using \eqref{eq:scalar_C}, one finds
\begin{equation}\label{eq:Dilatino_condition_chiral_projection}
  0 = D_R t_I^{\phantom{I}J} \xi_J + t_I^{\phantom{I}J} R^l (\mathcal{F} + 2 \mathcal{V})^{mn} \Gamma_{lmn} \xi_J + s^{-1} (\mathcal{F} + 2 \mathcal{V})^{mn} (t^{KL}\Theta_{KL})_{mn} \xi_I.
\end{equation}
A related identity can be obtained by contracting \eqref{eq:Dilatino} with $\xi^I \Gamma_{mn}$ and projecting onto the horizontal subspace:
\begin{equation}\label{eq:D_R_t_IJ-identity}
  \left( {{R^k}{D_k}{t^{IJ}}} \right){\left( {{\Theta _{IJ}}} \right)_{mn}} - 2\left[ (\mathcal{F} + 2 \mathcal{V})^H \times (t^{IJ} \Theta_{IJ}) \right]_{mn} = 0.
\end{equation}
where $(\eta \times \omega)_{mn} = \eta_m^{\phantom{m}p} \omega_{pn} - \omega_m^{\phantom{m}p} \eta_{pn}$. In passing, one needs to use the simple identity \eqref{vanishing-product}. As a point of consistency note that one can obtain the same result by contracting \eqref{eq:Dilatino_condition_chiral_projection} with $\xi^I \Gamma_{mn}$ and again projecting onto the horizontal part.

\subsection{Integrability}

\subsubsection{Cauchy-Riemann structures}
\label{sec:CR_integrability}

Having established the existence of the almost CR structure \eqref{eq:almost-CR-structure}, it is natural to ask if it satisfies any integrability condition. As a warm-up to the integrability condition of a THF \eqref{eq:THF_integrability_condition}, we consider the slightly simpler case of a CR structure \eqref{eq:CR_integrability_condition}.

Thus we study the condition \eqref{spinorial-(1,0)} for the commutator $[X,Y]$ for arbitrary $X$, $Y \in T^{1,0}$. I.e.~by acting with $Y^n D_n$ on \eqref{spinorial-(1,0)} and antisymmetrizing in $X, Y$, one finds that
\begin{equation}
  [X, Y] \in T^{1,0} 
  \quad\Leftrightarrow\quad
  0 = X^{[m} Y^{n]} \left[ D_m H_I^{\phantom{I}J} \Gamma_n \xi_J + H_I^{\phantom{I}J} \Gamma_n D_m \xi_J \right].
\end{equation}
This reduces quickly to
\begin{equation}\label{eq:CR_integrability_intermediate_eq}
  X^{[m} Y^{n]} \left[ D_m H_I^{\phantom{I}J} \Gamma_n \xi_J - [H, t]_I^{\phantom{I}J} \Gamma_{mn} \xi_J + 2 H_I^{\phantom{I}J} (\mathcal{F} + \mathcal{V})_{mn} \xi_J \right].
\end{equation}
Per usual, \eqref{eq:CR_integrability_intermediate_eq} can be mapped to two equations by suitable contractions.

To begin, we contract \eqref{eq:CR_integrability_intermediate_eq} with $\xi^I$ and find that $( [H,t]_I^{\phantom{I}J} \Theta_J^{\phantom{J}I} )^{2,0} = 0$. Due to the reality conditions for $\xi_I$, $\mathfrak{m}_I^{\phantom{I}J}$ and $t_I^{\phantom{I}J}$ this means that $[H,t]_I^{\phantom{I}J} \Theta_J^{\phantom{J}I} \in \Omega^{1,1}$. This in turn is equivalent to $[H, t]_I^{\phantom{I}J}$ being proportional to $\mathfrak{m}_I^{\phantom{I}J}$. However, $[H, t]_I^{\phantom{I}J}$ is proportional to $[\mathfrak{m}, t]_I^{\phantom{I}J}$ and thus the only solution is $\mathfrak{m}_I^{\phantom{I}J} = f t_I^{\phantom{I}J}$ for any non-zero function $f$.

Being rid of the commutator term, we consider the contraction with $\xi_J$ symmetrized over $SU(2)_R$ indices. This leads to $s H_{IJ} [X^{[m} Y^{n]} (\mathcal{F} + \mathcal{V})_{mn}]$. The necessary vanishing of the expression in square brackets means that $(\mathcal{F} + \mathcal{V})^{2,0} = 0$.

Finally, we contract with $\xi_J \Gamma_k$:
\begin{equation}
  X^{[m} Y^{n]} D_m H_I^{\phantom{I}K} \left(\Theta_{JKkn} - \frac{1}{2} \epsilon_{JK} s g_{kn}\right).
\end{equation}
By symmetrizing and antisymmetrizing over $I$ and $J$, it is clear that both terms in parantheses have to vanish independently. It follows that $D_X H_I^{\phantom{I}J} = 0$.

In summary, the almost CR-structure is integrable and the manifold is CR if and only if
\begin{equation}\label{eq:CR_integrability_obstructions}
  \mathfrak{m}_I^{\phantom{I}J} = f t_I^{\phantom{I}J}, \qquad
  (\mathcal{F} + \mathcal{V})^{2,0} = 0, \qquad
  D_X \left(\frac{t_I^{\phantom{I}J}}{\sqrt{\det t}}\right) = 0, \qquad
  \forall X \in T^{1,0}.
\end{equation}
Note that due to our reality condition for $t_I^{\phantom{I}J}$, the last statement is actually equivalent to
\begin{equation}
  D_X \left(\frac{t_I^{\phantom{I}J}}{\sqrt{\det t}}\right) = 0, \qquad
  \forall X \in TM_H,
\end{equation}
where $TM_H = T^{1,0} \oplus T^{0,1}$.

\subsubsection{Transversally holomorphic foliations}

Having discussed integrable CR structures, we now turn to the integrability condition for transversal holomorphic foliations \eqref{eq:THF_integrability_condition}. Using identical arguments to those from the previous section, we note that the integrability condition is
\begin{equation}\label{eq:spinorial_THF_integrability_condition}
  [X, Y] \in T^{1,0} \oplus R
  \quad\Leftrightarrow\quad
  0 = X^{[m} Y^{n]} \left[ D_m H_I^{\phantom{I}J} \Pi_n^{\phantom{n}k} \Gamma_k \xi_J + \nabla_m \Pi_n^{\phantom{n}k} H_I^{\phantom{I}J} \Gamma_k \xi_J + H_I^{\phantom{I}J} \Pi_n^{\phantom{n}k} \Gamma_k D_m \xi_J \right].
\end{equation}

To begin, consider \eqref{eq:spinorial_THF_integrability_condition} for $X, Y \in T^{1,0}$. Direct substitution gives
\begin{equation}
  X^{[m} Y^{n]} ( D_m H_I^{\phantom{I}J} \Gamma_n - [H, t]_I^{\phantom{I}J} \Gamma_{mn} -2 \star \mathcal{V}_{mnk} (\Gamma^k + R^k) H_I^{\phantom{I}J} ) \xi_J
\end{equation}
Now, since $X, Y \in T^{1,0}$, the only contributions to the last term arise from the components of $\star \mathcal{V}$ that lie in $\Omega^{2,1} \oplus \Omega^{2,0} \wedge R$. However, since $(\Gamma^k + R^k) \xi_I = \Pi^k_{\phantom{k}l} \Gamma^l \xi_I$ the latter of these is annihilated by the projection while the former vanishes due to holomorphy --- i.e.~ for any $\omega \in \Omega^{0,1}$, $H_I^{\phantom{I}J} \omega_k \Gamma^k \xi_J = 0$. Thus we are left with
\begin{equation}
  X^{[m} Y^{n]} ( D_m H_I^{\phantom{I}J} \Gamma_n - [H, t]_I^{\phantom{I}J} \Gamma_{mn}) \xi_J
\end{equation}
Once again, contraction with $\xi^I$ gives the first necessary condition, $([H, t]^{IJ} \Theta_{IJ})^{2,0} = 0$, from which it follows once again that $\mathfrak{m}_I^{\phantom{I}J} = f t_I^{\phantom{I}J}$. Just as in the CR case the second condition is $D_X H_I^{\phantom{I}J} = 0$, $\forall X \in T^{1,0}$.

We continue our analysis of \eqref{eq:spinorial_THF_integrability_condition} by considering $X \in T^{1,0}$ and $Y = R$. Using the results from the previous paragraph, one finds that the necessary and sufficient condition is the vanishing of
\begin{equation}\label{eq:THF_integrability_final_piece}
  X^m [ - D_R H_I^{\phantom{I}J} \Gamma_m + 2 (\mathcal{F} + 2 \iota_R \star \mathcal{V})_{mn} (\Gamma^n + R^n) H_I^{\phantom{I}J} ] \xi_J.
\end{equation}
By inspection one finds that the only contributing terms including $\mathcal{F}$ or $\mathcal{V}$ lie in $\Omega^{2,0}$ --- $(\Omega^{1,0} \oplus \Omega^{0,1}) \wedge R$ as well as $\Omega^{0,2}$ components are projected to zero while those in $\Omega^{1,1}$ vanish due to holomorphy. The components in $\Omega^{2,0}$ are of course self-dual under $\iota_R \star$ so the above can be rewritten in terms of $\mathcal{F} + 2 \mathcal{V}$ instead of $\mathcal{F} + 2 \iota_R \star \mathcal{V}$.

To further simplify this, we consider the chiral projection of the dilatino equation \eqref{eq:Dilatino_condition_chiral_projection}. Acting with $X^m H_I^{\phantom{I}J} \Gamma_m$ on \eqref{eq:Dilatino_condition_chiral_projection} one finds that
\begin{equation}\label{eq:THF_integrability_another_auxiliary}
  H_I^{\phantom{I}J} D_R t_J^{\phantom{J}K} X^m \Gamma_m \xi_K = 4 H_I^{\phantom{I}J} t_J^{\phantom{J}K} X^m \iota_R \star (\mathcal{F} + 2 \mathcal{V})_{mn} \Gamma^n \xi_K.
\end{equation}
Now, we first note that $D_R t_I^{\phantom{I}J} = \sqrt{\det t} D_R H_I^{\phantom{I}J}$ as $D_R (\det t) = 0$. Together with $H_I^{\phantom{I}K} H_K^{\phantom{K}J} = - 2 \imath H_I^{\phantom{I}J}$ it follows that
\begin{equation}
  D_R H_I^{\phantom{I}J} X^m \Gamma_m \xi_K = 2 \imath (\det t)^{-1/2} X^m H_I^{\phantom{I}J} t_J^{\phantom{J}K} \iota_R \star (\mathcal{F} + 2 \mathcal{V})_{mn} \Gamma^n \xi_K = 2 X^m H_I^{\phantom{I}J} \iota_R \star (\mathcal{F} + 2\mathcal{V})_{mn} \Gamma^n \xi_K.
\end{equation}
As before we argue that only the $\Omega^{2,0}$ and $\Omega^{0,2}$ terms contribute. Thus we find that \eqref{eq:THF_integrability_final_piece} vanishes without any further conditions. In the end, the integrability conditions are
\begin{equation}\label{eq:THF_integrability_obstructions}
  \mathfrak{m}_I^{\phantom{I}J} = f t_I^{\phantom{I}J}, \qquad
  D_X \left(\frac{t_I^{\phantom{I}J}}{\sqrt{\det t}}\right) = 0, \qquad
  \forall X \in T^{1,0}.
\end{equation}
As in the case of the CR structure the reality condition for $t_I^{\phantom{I}J}$ implies that the last condition holds for all horizontal sections of the tangent bundle. By comparison with equation \eqref{eq:CR_integrability_obstructions} it is clear that any solution defining a THF also defines an integrable CR structure while the converse is not the case.

\section{Implications for Localization}
\label{sec:implications_for_localization}

\subsection{The $\partial_b$ and $\bar \partial_b$ operators}

Suppose that our manifold satisfies either of the integrability conditions \eqref{eq:CR_integrability_obstructions} or \eqref{eq:THF_integrability_obstructions}. Let us show one can define nilponent operators $\partial_b$ and $\bar \partial_b$ similar to those on complex structures. To do so, consider a $(0,1)$-form ${\alpha ^{0,1}}$. We can decompose its exterior derivative as
\begin{equation}
  d{\alpha ^{0,1}} = {\pi _V}\left( {d{\alpha ^{0,1}}} \right) + {\pi ^{2,0}}\left( {d{\alpha ^{0,1}}} \right) + {\pi ^{1,1}}\left( {d{\alpha ^{0,1}}} \right) + {\pi ^{0,2}}\left( {d{\alpha ^{0,1}}} \right),
\end{equation}
where $\pi_V$ and $\pi^{p,q}$ are projectors to the vertical and $(p,q)$ components. Since neither $[T^{1,0} \oplus \bC R, T^{1,0} \oplus \bC R]$ nor $[T^{1,0}, T^{1,0}]$ have a component in $T^{0,1}$ one finds that
  \begin{equation}
    d{\alpha ^{0,1}}\left( {{X^{1,0}},{Y^{1,0}}} \right) = {X^{1,0}}\left( {{\alpha ^{0,1}}\left( {{Y^{1,0}}} \right)} \right) - {Y^{1,0}}\left( {{\alpha ^{0,1}}\left( {{X^{1,0}}} \right)} \right) - {\alpha ^{0,1}}\left( {\left[ {{X^{1,0}},{Y^{1,0}}} \right]} \right).
  \end{equation}
In other words, ${\pi ^{2,0}}\left( {d{\alpha ^{0,1}}} \right) = 0$, which allows us to define $(d_V, \partial_b, \bar{\partial}_b)$ via
  \begin{equation}
    d{\alpha ^{0,1}} = {\pi _V}\left( {d{\alpha ^{0,1}}} \right) + {\pi ^{1,1}}\left( {d{\alpha ^{0,1}}} \right) + {\pi ^{0,2}}\left( {d{\alpha ^{0,1}}} \right) \equiv {d_V}{\alpha ^{0,1}} + \partial_b {\alpha ^{0,1}} + \bar{\partial}_b {\alpha ^{0,1}}.
  \end{equation}
From $d = \partial_b + \bar{\partial}_b + d_v$ and $d^2 = 0$ it follows directly that $\partial_b^2 = \bar{\partial}_b^2 = 0$ and one can define cohomology groups $H^{p,q}_{\bar{\partial}_b}$ via the exact sequence
\begin{equation}
  \dots \xrightarrow{\bar{\partial}_b} \Omega^{p,q-1} \xrightarrow{\bar{\partial}_b} \Omega^{p,q} \xrightarrow{\bar{\partial}_b} \Omega^{p,q+1} \xrightarrow{\bar{\partial}_b} \dots .
\end{equation}

\subsection{Mode counting and partition functions}

As mentioned in the introduction, partition functions for supersymmetric gauge theories calculated in the context of topological field theories or localization simplify significantly on K\"ahler and Sasakian manifolds. The argument relies not only on the existence of the differential $\bar{\partial}_b$ ($\bar{\partial}$ in the K\"ahler case). Indeed, one also requires the compatibility of the decomposition \eqref{eq:almost-CR-structure} with the action of the Lie derivative $\pounds_{sR}$. In this section we will go over this argument of \cite{Schmude:2014lfa,Rodriguez-Gomez:2014eza} in some detail and discuss under what circumstances it applies to the manifolds in question.

Consider a vector multiplet with Lie algebra $\mathfrak{g}$. The bosonic modes lie in $\Omega^1(\mathfrak{g}) \oplus H^0(\mathfrak{g}) \oplus H^0(\mathfrak{g})$, where $H^0(\mathfrak{g})$ denotes harmonic Lie algebra valued functions. Fermionic modes on the other hand can be mapped to $\Omega^+(\mathfrak{g}) \oplus \Omega^0(\mathfrak{g}) \oplus \Omega^0(\mathfrak{g})$. The one-loop contribution to the perturbative partition function is given by\footnote{This was shown to be true for generic Sasakian manifolds in \cite{Qiu:2013pta}. Here we assume it to be true for five-dimensional Riemannian manifolds admitting a integrable CR-structure or THF.}
\begin{equation}\label{eq:localization_determinant}
  \sqrt{\frac{\det_{\text{fermions}} \pounds_{sR}}{\det_{\text{bosons}} \pounds_{sR}}}.
\end{equation}
If $\pounds_{sR} \Phi = \pounds_{sR} \kappa = 0$ we can calculate the determinants using the decomposition \eqref{eq:almost-CR-structure}. Clearly $\pounds_{sR} \kappa = 0$, so we need to evaluate $\pounds_{sR} \Phi = \iota_{sR} d\Phi$. Direct calculation using \eqref{eq:Killing-3} yields
\begin{equation}
  d\Phi = - s^{-1} ds \wedge \Phi + s^{-1} D \left(\frac{t^{IJ}}{\sqrt{\det t}}\right) \wedge \Theta_{IJ} + 2 s^{-1} \left[ \iota_R (\mathcal{F} + 2 \mathcal{V}) \wedge \Phi - \kappa \wedge ((\mathcal{F} + 2 \mathcal{V}) \times \Phi) \right],
\end{equation}
Thus
\begin{equation}
  \pounds_{sR} \Phi = D_R \left(\frac{t^{IJ}}{\sqrt{\det t}}\right) \Theta_{IJ} - 2 \left[ (\mathcal{F} + 2 \mathcal{V})^H \times \Phi \right] = 0
\end{equation}
where we used \eqref{eq:D_R_t_IJ-identity}. In conclusion we can rewrite \eqref{eq:localization_determinant} as
\begin{equation}
  \sqrt{\frac{\det_{\pounds_{sR}} (\Omega^{2,0} \oplus \Omega^{0,0} \Phi \oplus \Omega^{0,2} \oplus \Omega^{0,0} \oplus \Omega^{0,0})}{\det_{\pounds_{sR}}(\Omega^{1,0} \oplus \Omega^{0,1} \oplus \Omega^{0,0} \kappa)}} \frac{1}{\det_{\pounds_{sR}} H^0},
\end{equation}
where we used the notation $\det_{\Omega^{p,q}} \pounds_{sR} = \det_{\pounds_{sR}} \Omega^{p,q}$ and dropped the various appearances of $\mathfrak{g}$ for readability. As $[\pounds_{sR}, \bar{\partial}_b] = 0$ it follows that the above simplifies to
\begin{equation}
  \sqrt{\frac{\det_{\pounds_{sR}} H^{0,2}_{\bar{\partial}_b} \det_{\pounds_{sR}} H^{0,0}_{\bar{\partial}_b}}{\det_{\pounds_{sR}} H^{0,1}_{\bar{\partial}_b}}}
  \sqrt{\frac{\det_{\pounds_{sR}} H^{2,0}_{\bar{\partial}_b} \det_{\pounds_{sR}} H^{0,0}_{\bar{\partial}_b}}{\det_{\pounds_{sR}} H^{1,0}_{\bar{\partial}_b}}}.
\end{equation}

It is interesting to note that the above argument does not require a property akin to Lefschetz decomposition on K\"ahler manifolds. Recall that the Lefschetz theorem relates cohomology groups of the Dolbeault operator as $H^{0,0}_{\bar{\partial}} \cong H^{1,1}_{\bar{\partial}, \omega}$, where the subscript $\omega$ denotes forms parallel to the symplectic form $\omega$. Such a decomposition, while true for e.g.~Sasaki-Einstein manifolds does not hold in general for the operator $\Phi$. That is, for $\alpha \in \Omega^{1,1}_\Phi$ one can write $\alpha = a \Phi$ for some scalar function $a$, yet $\bar{\partial}_b \alpha = 0$ is not in one-to-one correspondence with $\bar{\partial}_b a = 0$ since $\bar{\partial}_b \Phi$ does not vanish in the general case.

\subsection{BPS equations on the Higgs branch}

The nilponency of $\bar{\partial}_b$ has also immediate implications on the Higgs branch BPS equations of $\mathcal{N} = 1$ theories. In \cite{Pan:2014bwa} these were studied for supersymmetric backgrounds that are K-contact. Defining $\bar{\partial}_a \equiv \bar{\partial}_b - \imath a^{0,1}$ for a $U(1)$ connection $a$ with field strength $F_a$, some of the relevant equations are
\begin{equation}
  {\bar \partial _a}\alpha  + \bar \partial _a^*\beta  = 0,
  \qquad
  F_a^{0,2} = 2i\bar \alpha \beta ,
  \qquad
  F_a^{d\kappa } = \frac{1}{2}\left( {\zeta  - {{\left| \alpha  \right|}^2} + {{\left| \beta  \right|}^2}} \right).
\end{equation}
Here, $\alpha$ is a 0-form and $\beta$ is a $(0,2)$-form; both are related to the scalar in the hypermultiplet. The superscript $d\kappa$ denotes the component along $d\kappa$. The BPS equations and the nilpotence then imply that $\bar{\partial}_a \bar{\partial}_a^* \beta = - \bar{\partial}_a \bar{\partial}_a \alpha = \imath F_a^{0,2} \alpha =  -2 \vert \alpha \vert^2 \beta$. Thus $2\int \vert \alpha \vert^2 \vert \beta \vert^2 + \int \vert \bar{\partial}_a^* \beta \vert^2 = 0$, and it follows that
    \begin{equation}
      \left\{ \begin{gathered}
      \beta  = 0 \hfill \\
      {{\bar \partial }_a}\alpha  = 0 \hfill \\ 
      \end{gathered}  \right.,
      \quad
      {\text{or}}
      \quad
      \left\{ \begin{gathered}
      \alpha  = 0 \hfill \\
      \bar \partial _a^*\beta  = 0 \hfill \\ 
      \end{gathered}  \right.
      .
    \end{equation}
In other words, similar to our discussion in the previous section we see that results for Sasaki (-Einstein) manifolds can be extended to geometries that are either THF or CR.

\section{A Karlhede-Rocek-Witten twist in five dimensions}
\label{sec:krw_twist}

As discussed above in section \ref{sec:susy_variations_and_bispinors} as well as in \cite{Imamura:2014ima} a necessary condition for the existence of a solution of the background supergravity variations for supersymmetry spinors satisfying the symplectic Majorana condition is the existence of a Killing vector. Recall that the symplectic Majorana condition \eqref{eq:symplectic_Majorana_condition} implies that $s > 0$ from which it follows that $v$ has no zeroes. In other words, the Killing vector is globally non-vanishing.\footnote{If one does not impose the symplectic Majorana condition, the situation is more complicated. I.e.~both $s$ and $R$ are generally complex; it is also clear that the vector vanishes if the spinors are parallel. Moreover, note that $R$ does not even vanish at a single point. Assume $\exists p \in \mathcal{M}$ such that $R\vert_p = 0$. It follows that $s(p) = 0$ and thus $\xi_I\vert_p = 0$. From the gravitino equation it follows immediately that $\xi_I$ vanishes identically on $\mathcal{M}$.}
In this section we will show that the existence of a globally non-vanishing Killing vector is also sufficient for the manifold $\mathcal{M}$ to admit supersymmetry spinors that solve \eqref{Killing} and \eqref{eq:Dilatino}.\footnote{We would like to thank Diego Rodriguez-Gomez for many discussions and collaboration that lead to the approach used in this section.}
At the heart of the argument is the idea that the existence of the vector implies that the manifold supports an $SO(4)$ structure. This in turn allows us to do a standard Witten twist \cite{Witten:1988ze,Karlhede:1988ax}.\footnote{See also e.g.~\cite{Witten:2011zz,Anderson:2012ck} for five-dimensional, twisted field theories.}
Our strategy is to work in a patch using methods familiar from Kaluza-Klein reduction, yet show that we can write the overall result in terms of globally well-defined objects. In principal one should be able to make the same argument using the general, local solution of \cite{Imamura:2014ima}.

Given a manifold $\mathcal{M}$ with a Killing vector $\mathfrak{v} = \partial_\tau$ we can write the vielbein as\footnote{
  In this section, greek indices run from one to five while roman ones only run from one to four.
}
\begin{equation}
  \hat{e}_\mu^{\phantom{\mu}\alpha} = \begin{pmatrix}
    e_m^{\phantom{m}a} & k a_m \\ 0 & k
  \end{pmatrix}, \qquad
  \hat{E}_\alpha^{\phantom{\alpha}\mu} = \begin{pmatrix}
    E_a^{\phantom{a}m} & - a_a \\ 0 & k^{-1}
  \end{pmatrix}.
\end{equation}
I.e.~the metric takes the form $ds^2 = g_{mn} dx^m dx^n + k^2 (d\tau + a)^2$, where $\partial_\tau = \mathfrak{v}$. The spin connection is
\begin{equation}
  \hat{\omega}_{abc} = \omega_{abc}, \quad
  \hat{\omega}_{ab5} = \frac{1}{2} k f_{ab}, \quad
  \hat{\omega}_{5bc} = - \frac{1}{2} k f_{bc}, \quad
  \hat{\omega}_{5b5} = - \partial_b \log k.
\end{equation}
Here, $f = da$. Keeping in mind \eqref{eq:bi-spinors_algebraic_properties}, we demand the spinor $\xi_I$ to be anti-chiral. That is, $\Gamma^5 \xi_I = - \xi_I$ which is why we write $\xi_I \equiv \xi_I^-$.

\subsection{Gravitino Equation}

One can then decompose the gravitino variation \eqref{Killing} into components along $a = 1, \dots 4$, components along $a = 5$ as well as chiral and anti-chiral parts:
\begin{IEEEeqnarray}{rCl}
  0 &=& D_a \xi_I^- - a_a (\partial_\tau \xi_I^- - A_{\tau I}^{\phantom{\tau I}J} \xi_J^-) + \mathcal{F}_{a5} \xi_I^- + \Gamma_{ab} \mathcal{V}^{b5} \xi_I^-, \label{eq:krw_twist_gravitino_i} \\
  0 &=& - \frac{1}{4} k f_{ab} \Gamma^b \xi_I^- - t_I^{\phantom{I}J} \Gamma_a \xi_J^- - \mathcal{F}_{ab} \Gamma^b \xi_I^- - \frac{1}{2} \mathcal{V}^{bc} \Gamma_{abc} \xi_I^-, \label{eq:krw_twist_gravitino_ii} \\
  0 &=& k^{-1} (\partial_\tau \xi_I^- - A_{\tau I}^{\phantom{\tau I}J} \xi_J^-) + t_I^{\phantom{I}J} \xi_J^- - \frac{1}{8} k f_{bc} \Gamma^{bc} \xi_I^- + \frac{1}{2} \mathcal{V}^{bc} \Gamma_{bc} \xi_I^-, \label{eq:krw_twist_gravitino_iii} \\
  0 &=& \frac{1}{2} \partial_b \log k \Gamma^b \xi_I^- + \mathcal{F}_{b5} \Gamma^b \xi_I^-. \label{eq:krw_twist_gravitino_iv}
\end{IEEEeqnarray}

The last of these, \eqref{eq:krw_twist_gravitino_iv}, is solved by $\mathcal{A} = - \frac{1}{2} k^{-1} \mathfrak{v}$. It follows that
\begin{equation}
  \mathcal{F}_{a5} = -\frac{1}{2} \partial_a \log k, \qquad
  \mathcal{F}_{ab} = -\frac{1}{2} k f_{ab}. 
\end{equation}
Equation \eqref{eq:krw_twist_gravitino_i} is solved by setting $A_{\tau I}^{\phantom{\tau I}J} = 0$, $\xi_I^- = \sqrt{k} \chi_I$, where $\chi_I \chi^I = 1$, and --- more importantly --- $D_a \chi_I = \nabla_a \chi_I - A_{aI}^{\phantom{aI}J} \chi_J = 0$. The possibility of finding a $\chi$ such that $\nabla_a \chi_I = A_{aI}^{\phantom{aI}J} \chi_J$ is of course at the heart of this argument. As long as $\Gamma^5 \chi_I = - \chi_I$, it is possible to find such a spinor; explicit calculations can be done using 't Hooft matrices for example \cite{Rodriguez-Gomez:2015xwa}. With all our previous assumptions and observations \eqref{eq:krw_twist_gravitino_ii} becomes
\begin{equation}
  4 \mathcal{V}_{ab} = \frac{1}{2} k \epsilon_{abcd5} f^{cd} + 4 s^{-1} \Theta_{ab}^{IJ} t_{IJ}.
\end{equation}
Substituting this into \eqref{eq:krw_twist_gravitino_iii} we find that $t_{IJ} = 0$ since
\begin{equation}
  0 = \frac{1}{8} \left( k f_{ab} - 4 \mathcal{V}_{ab} \right) \Theta_{IJ}^{ab} = \frac{1}{8} \left( k f_{ab} - \frac{1}{2} k \epsilon_{abcd5} f^{cd} + 4 s^{-1} \Theta_{ab}^{KL} t_{KL} \right) \Theta^{ab}_{IJ} = \frac{1}{2} \Theta_{IJ}^{ab} \Theta_{ab}^{KL} t_{KL}.
\end{equation}
In summary, the gravitino equation is fully solved by
\begin{equation}\label{eq:krw_twist_gravitino_solution}
    \xi_I^- = \sqrt{k} \chi_I, \quad
    D_a \chi_I = 0, \quad
    \mathcal{A} = -\frac{1}{2} k^{-1} \mathfrak{v}, \quad
    t_{IJ} = 0, \quad
    4 \mathcal{V}_{ab} = \frac{1}{2} k \epsilon_{abcd5} f^{cd}, \quad
    \mathcal{V}_{a5} = 0.
\end{equation}
By now it is clear that the spinor bilinears $s, v$ coincide with the scalar and vector defined by the background, $k, \mathfrak{v}$, i.e.~$s = k$, $v = \mathfrak{v}$, so we drop the distinction.

\subsection{Dilatino Equation}

Performing a similar decomposition of the Dilatino equation, one finds
\begin{IEEEeqnarray}{rCl}
  0 &=& 4 \hat{D} t_I^{\phantom{I}J} \Gamma^a \xi_J^- - 8 t_I^{\phantom{I}J} (\mathcal{F} + 2 \mathcal{V})_{a5} \Gamma^a \xi_J^- + 4 \hat{\nabla}_\alpha \mathcal{V}^{\alpha b} \Gamma_b \xi_I^- - 4 \mathcal{F}_{ab} \mathcal{F}_{c5} \Gamma^{abc} \xi_I^-, \label{eq:krw_twist_dilatino_i} \\
  0 &=& -4 \hat{D}_5 t_I^{\phantom{I}J} \xi_J^- + 4 t_I^{\phantom{I}J} (\mathcal{F} + 2 \mathcal{V})_{ab} \Gamma^{ab} \xi_I^- - 4 \hat{\nabla}_\alpha \mathcal{V}^{\alpha 5} \xi_I^- + \mathcal{F}_{ab} \mathcal{F}_{cd} \Gamma^{abcd} \xi_I^- + C \xi_I^-. \label{eq:krw_twist_dilatino_ii}
\end{IEEEeqnarray}
Imposing the solution to the gravitino equations \eqref{eq:krw_twist_gravitino_solution}, this simplifies of course considerably. Also, note that
\begin{equation}
  \hat{\nabla}_\alpha \mathcal{V}^{\alpha b} = \nabla_a \mathcal{V}^{ab} + \partial^a \log k \mathcal{V}_{ab}, \qquad
  \hat{\nabla}_\alpha \mathcal{V}^{\alpha 5} = -\frac{1}{2} k f_{ab} \mathcal{V}^{ab}.
\end{equation}
Then, \eqref{eq:krw_twist_dilatino_ii} is solved by $C = -\frac{1}{4} k^2 f_{ab} f_{cd} \epsilon^{abcd5}$. Since $4 s \nabla_b \mathcal{V}^{ba} = - \frac{1}{2} s \epsilon^{abcd5} f_{bc} \partial_d k$, one finds that \eqref{eq:krw_twist_dilatino_i} is solved trivially.

\subsection{Topological Issues}

To conclude, we discuss whether the solution \eqref{eq:krw_twist_gravitino_solution} is globally well-defined. Since $\mathcal{F}$ is globally exact we only have to worry about the $SU(2)_R$ field strength. Our strategy is to rewrite this in terms of the Riemann tensor. Thus we use the integrability condition for the spinor $\chi_I$,
\begin{equation}
  0 = [D_a, D_b] \chi_I = - F^{IJ}_{ab} \chi_J + \frac{1}{4} R_{ab\alpha\beta} \gamma^{\alpha\beta} \chi_I.
\end{equation}
This implies $F_{ab}^{IJ} = -2 s^{-1} R_{ab\kappa\lambda} \Theta^{IJ\kappa\lambda}$ from which it follows that we can express the $SU(2)_R$ connection in terms of a projection of the Riemann tensor. In summary, the two connections are
\begin{equation}\label{eq:krw_twist_fluxes}
  F^{IJ} = -2 s^{-1} R^H_{\mu\nu} \Theta^{IJ\mu\nu}, \qquad
  \mathcal{F} = -\frac{1}{2} d (k^{-1}\mathfrak{v}).
\end{equation}
where $R^H_{\mu\nu} = \Pi_\kappa^\sigma \Pi_\lambda^\tau R_{\sigma\tau\mu\nu} dx^\kappa \otimes dx^\lambda$ denotes the horizontal part of the curvature two-form. In both cases, all objects appearing on the right hand side are globally well defined. We proceed to consider characteristic classes defined by $F^{IJ}$. Using \eqref{eq:involved_theta_identities} one finds that
\begin{equation}\label{eq:characteristic_class_of_SU2_R}
  F_I^{\phantom{I}J} \wedge F_J^{\phantom{J}I} = - 4 R^H_{\kappa\lambda} \wedge R^H_{\mu\nu} \left( \Pi^{\kappa\mu} \Pi^{\lambda\nu} + \frac{1}{2} \epsilon^{\kappa\lambda\mu\nu\rho} \kappa_\rho \right).
\end{equation}
The expression is completely horizontal and since $\mathfrak{v}$ is Killing, $0 = \pounds_{\mathfrak{v}} (F_I^{\phantom{I}J} \wedge F_J^{\phantom{J}I}) = \iota_{\mathfrak{v}} d (F_I^{\phantom{I}J} \wedge F_J^{\phantom{J}I})$ from which it follows that \eqref{eq:characteristic_class_of_SU2_R} is closed and defines thus an element of the de Rham cohomolgy group $H^4(\mathcal{M})$ as it should. Usually the next question would be whether this element is trivial and whether it might be an obstruction to the existence of the solution given by \eqref{eq:krw_twist_gravitino_solution} and $C$. However, equation \eqref{eq:characteristic_class_of_SU2_R} clearly show that this class has a representative that is independent of our specific solution since it can be expressed in terms of $\mathfrak{v}$ and the Riemann tensor. Thus, in the case that the class is non-trivial, it is clear that the corresponding cycle in homology exists and vice versa.

One might worry about the $f$ dependence of $\mathcal{V}$. In general, the manifolds are not bundles yet only foliations and one cannot necessarily think of $f$ as the curvature of a connection. Yet as we saw above, $f$ is a projection of $\mathcal{F}$ onto the horizontal space --- $f = -2 k^{-1} \mathcal{F}^H$. While one might not consider $f$ globally as the curvature of a connection, it is well-defined as a two-form. Since it doesn't enter the solution directly yet only via $\mathcal{V}$, this is good enough and we conclude that any manifold $\mathcal{M}$ admits a solution to \eqref{Killing} and \eqref{eq:Dilatino} with symplectic Majorana spinor if and only if there is a non-vanishing Killing vector $\mathfrak{v}$.

\section{Examples}
\label{sec:examples}

It follows from the previous section that any direct product $\mathbb{R} \times \mathcal{M}_4$ or $S^1 \times \mathcal{M}_4$ admits a solution to the Killing spinor equations and thus rigid supersymmetry. Similarly, it is clear that such manifolds do at least not trivially\footnote{
  ``Trivially'' here means that one simply embeds the Killing vector in the obvious way. For a specific choice of $\mathcal{M}_4$ and Killing vector, this might change.
}
admit an integrable CR-structure or a THF if $\mathcal{M}_4$ does not admit a complex structure --- the example coming to mind here being $\mathbb{R} \times S^4$. See however the discussion in \cite{Alday:2015lta}.

\subsection{Sasakian manifolds}

Sasakian manifolds are the odd-dimensional analogues of K\"ahler manifolds. They are either characterized by having K\"ahler metric cones, or by the existence of a Killing spinor satisfying
  \begin{equation}
    \left( {{\nabla _m} - i{\mathcal{A}_m}} \right)\xi  = \frac{i}{2}{\Gamma _m}\xi.
  \end{equation}
Here, $\mathcal{A}$ is the connection one-form associated to the Ricci-form on the metric cone. The equation and its complex conjugate correspond to the special case of \eqref{Killing} with
  \begin{equation}
    \mathcal{F} = \mathcal{V} = 0,
    \qquad
    {\left( {{\mathcal{A}_m}} \right)_I}^J = {\mathcal{A}_m}{\left( {{\sigma _3}} \right)_I}^J,
    \qquad
    {t_I}^J = \frac{i}{2}{\left( {{\sigma _3}} \right)_I}^J.
  \end{equation}
Since both $t$ and $\mathcal{A}$ have only components along $\sigma_3$ one finds that $\nabla_m t_{IJ} = 0$. The dilatino equation is solved by
  \begin{equation}
    C = 0.
  \end{equation}
Hence, $\mathcal{N} = 1$ supersymmetry can be defined on any 5-dimensional Sasakian structure as was first observed without resorting to supergravity \cite{Qiu:2013pta}.

Sasakian structures are examples of both Cauchy-Riemann or transversal-holomorphic structures, as follows from the fact that ${\nabla _m}{t_{IJ}} = \mathcal{F} = \mathcal{V} = 0$.

\subsection{Squashed $S^5$ with $SU(3) \times U(1)$ Symmetry}

Squashed five-spheres have appeared in various places in the literature. In particular, \cite{Imamura:2012bm,Alday:2014bta} discussed a class of squashed $S^5$, with the metric
  \begin{equation}\label{eq:squashed_s5_metric}
    ds_{S_b^5}^2 = \frac{1}{{{b^2}}}{\left( {d\tau  + h} \right)^2} + d{\sigma ^2} + \frac{1}{4}{\sin ^2}\sigma \left( {d{\theta ^2} + {{\sin }^2}\theta d{\varphi ^2}} \right) + \frac{1}{{16}}{\sin ^2}2\sigma {\left( {d\psi  + \cos \theta d\varphi } \right)^2}.
  \end{equation}
Our discussion follows that of \cite{Alday:2014bta} closely. The real constant $b$ is the squashing parameter, which gives a round sphere when $b = 1$, $h$ is a 1-form defined as
  \begin{equation}
    h =  - \frac{1}{2}{\sin ^2}\sigma \left( {d\psi  + \cos \theta d\varphi } \right).
  \end{equation}
where $\omega$ can be viewed as the K\"ahler form on $\mathbb{C}P^2$, satisfying $d\omega = 0$. The metric is written in a form adapted to the smooth $U(1)$-fibration over $\mathbb{C}P^2$, where ${b^{ - 2}}{\left( {d\tau  + h} \right)^2}$ is the metric in the $U(1)$-fiber direction, and $b$ is there to squash the radius. In this way it is easy to see the metric has $U(1)\times SU(3)$ symmetry, where $U(1)$ rotates the fiber, and $SU(3)$ is the isometry of $\mathbb{C}P^2$. The $\mathbb{C}P^2$ K\"ahler form is $\omega = \frac{1}{2} dh$. With the vielbein
\begin{equation}
  e^1 = \frac{1}{2} \sin\sigma \cos\sigma \tau_3, \qquad
  e^2 = d\sigma, \qquad
  e^3 = \frac{1}{2} \sin\sigma \tau_2, \qquad
  e^4 = \frac{1}{2} \sin\sigma \tau_1, \qquad
  e^5 = b^{-1} (d\tau + h),
\end{equation}
one finds
  \begin{equation}
    \omega  = e^1 \wedge e^2 - e^3 \wedge e^4,
    \qquad
    \omega  \wedge \omega  = -2 {e^1} \wedge {e^2} \wedge {e^3} \wedge {e^4},
    \qquad
    *\left( {\omega  \wedge \omega } \right) = -2 e^5,
  \end{equation}
where we have introduced the left-invariant one forms
\begin{equation}
  \tau_1 + \imath \tau_2 = e^{-\imath \psi} (d\theta + \imath \sin\theta d\phi), \qquad
  \tau_3 = d\psi + \cos\theta d\phi.
\end{equation}

This class of squashed sphere admits solutions to the Killing spinor equations
  \begin{equation}\label{eq:S5_Killing_Spinor_Equation}
    {\nabla _m}{\xi _I} + \frac{i}{2}{\left( {{\mathcal{A}_m}} \right)_I}^J \xi_J =  - \frac{i}{{2b}}\left( {1 + Q\sqrt {1 - {b^2}} } \right){\left( {{\sigma _3}} \right)_I}^J{\Gamma _m}{\xi _J} + \frac{{\sqrt {1 - {b^2}} }}{b}{\omega _{mn}}{\Gamma ^n}{\xi _I} + \frac{1}{2}\frac{{\sqrt {1 - {b^2}} }}{{2b}}{\omega ^{pq}}{\Gamma _{mpq}}{\xi _I}.
  \end{equation}
where $Q$ is a real parameter. And of course one can define bilinears as in \eqref{eq:bispinors}. In terms of \eqref{eq:Dirac_matrices}, the quarter BPS solution with $Q = -3$ is given by
\begin{equation}
  \xi_1 = \frac{c_+}{\sqrt{2}} e^{-\frac{3\imath \tau}{2}} \begin{pmatrix} 1 \\ 1 \\ 0 \\ 0 \end{pmatrix}, \qquad
  \xi_2 = \frac{c_-}{\sqrt{2}} e^{\frac{3\imath\tau}{2}} \begin{pmatrix} 1 \\ -1 \\ 0 \\ 0 \end{pmatrix}.
\end{equation}
The symplectic Majorana condition \eqref{eq:symplectic_Majorana_condition} corresponds to $(c_-)^* = c_+$. For more involved $3/4$ BPS solutions refer to \cite{Alday:2014bta}.

Comparing \eqref{Killing} with \eqref{eq:S5_Killing_Spinor_Equation} one identifies
  \begin{equation}\label{eq:squashed_s5_solution}
    {t_I}^J =  - \frac{i}{{2b}}\left( {1 + Q\sqrt {1 - {b^2}} } \right){\left( {{\sigma _3}} \right)_I}^J,
    \;\;\;\;
    \mathcal{F} = \frac{{\sqrt {1 - {b^2}} }}{b}\omega ,
    \;\;\;\;
    \mathcal{V} = \frac{{\sqrt {1 - {b^2}} }}{{2b}}\omega ,
    \;\;\;\;
    {\left( {{\mathcal{A}_m}} \right)_I}^J = \frac{{\left( {1 + Q\sqrt {1 - {b^2}} } \right)\sqrt {1 - {b^2}} }}{b}{e^5}.
  \end{equation}
Note that $\kappa = - e^5$ and one finds that $\omega$ is horizontal and self-dual since $\star \omega = \kappa \wedge \omega$. Furthermore $d\kappa = -2 b^{-1} \omega$ and ${\nabla ^m}{\omega _{mn}} = 4 b^{-1} \kappa_n$. Moreover
  \begin{equation}
    {\omega _{mn}}{\omega _{kl}}{\Gamma ^{mnkl}}{\xi _I} = {\omega _{mn}} {\omega _{kl}}{\epsilon ^{mnkl}}_r{\Gamma ^r}{\xi _I} = 2{\omega _{mn}}{\left( {*\omega } \right)^{mn}}_r{\Gamma ^r}{\xi _I} = 2{\omega _{mn}}{\omega ^{mn}} \kappa_r {\Gamma ^r}{\xi _I} = -8 \xi _I.
  \end{equation}
Finally, substituting everything into the dilatino equation (\ref{eq:Dilatino}), one finds
\begin{equation}
  0 = -\frac{4\imath}{b^2} (1 + Q \sqrt{1-b^2}) \sqrt{1-b^2} (\sigma_3)_I^{\phantom{I}J} \omega_{mn} \Gamma^{mn} \xi_J + 8 \frac{\sqrt{1-b^2}}{b^2} \kappa_m \Gamma^m \xi_I - 8 \frac{1-b^2}{b^2} \xi_I + C \xi_I.
\end{equation}
From $\eqref{eq:scalar_C}$ it follows that
\begin{equation}
  C = 8 \frac{\sqrt{1-b^2}}{b^2} + 8 \frac{1-b^2}{b^2} - 4\imath \frac{(1+Q\sqrt{1-b^2})\sqrt{1-b^2}}{b^2 s} \omega^{mn} \Theta_{mnI}^{\phantom{mnI}J} (\sigma_3)_J^{\phantom{J}I},
\end{equation}
so the above simplifies to
\begin{equation}\label{eq:Dilatino_equation_for_S5_substituted}
  0 =  - 4 \imath \frac{(1 + Q \sqrt{1-b^2}) \sqrt{1-b^2}}{b^2} \left[ (\sigma_3)_I^{\phantom{I}J} \omega_{mn} \Gamma^{mn} \xi_J + s^{-1} \omega^{mn} \Theta_{mnK}^{\phantom{mnK}L} (\sigma_3)_L^{\phantom{L}K} \xi_I \right],
\end{equation}
which vanishes identically for the above solution.

Now compare the ``algebraic equation'' of \cite{Alday:2014bta}. Rewritten in our conventions, it is
\begin{equation}\label{eq:AFGRS_algebraic_equation}
  0 = (1+Q) \sqrt{1-b^2} \xi_I - \frac{\imath}{2} \sqrt{1-b^2} (\sigma_3)_I^{\phantom{I}J} \omega_{mn} \Gamma^{mn} \xi_J,
\end{equation}
where we used \eqref{eq:bi-spinors_algebraic_properties}. Contracting with $\xi^I$ one finds $(\sigma_3)_{I}^{\phantom{I}J} \omega^{mn} \Theta_{mnJ}^{\phantom{mnJ}I} = 2 \imath s (1+Q)$. Substituting this into \eqref{eq:Dilatino_equation_for_S5_substituted} yields \eqref{eq:AFGRS_algebraic_equation}, which tells us that the Dilatino equation and the ``algebraic equation'' are equivalent in the case of squashed $S^5$.

Comparing \eqref{eq:squashed_s5_solution} with \eqref{eq:CR_integrability_obstructions} and \eqref{eq:THF_integrability_obstructions} it is clear that the squashing does not change the fact that $S^5$ admits both a CR-structure and a THF. In principle this is already clear from the form of the metric \eqref{eq:squashed_s5_metric} since changes in the parameter $b$ do not affect the $\mathbb{C}P^2$ base of the bundle.

\section*{Acknowledgements}

We would like to thank Yolanda Lozano, Dario Martelli, Patrick Meessen, Alessandro Pini, Paul Richmond, Martin Ro{\v c}ek, Diego Rodriguez-Gomez, James Sparks, Alessandro Tomasiello and Maxim Zabzine for various discussions. J.S.~would also like to thank Kavli IPMU, the University of Milano-Bicocca and the organizers of the 2014 workshop ``The String Theory Universe'' where some of the results here were presented. The work of J.S.~is funded by by the Asturian government's CLARIN grant ACB14-27. The work of Y.P.~was supported in part by National Science Foundation, grant PHY-1316617.

\appendix

\section{Conventions}

\paragraph{Gamma matrices and spinors:}

  Let us focus on a Riemannian five-manifold $M$.
  We use $\Gamma^m$ to denote the hermitian Gamma matrices satisfying the algebra $\left\{ {{\Gamma _m},{\Gamma _n}} \right\} = 2{g_{mn}}$ for $g$ with the Euclidean signature. The spinors have 4 complex components. Denote the antisymmetric charge conjugation matrix by $C$, which satisfies $C{\Gamma _m} = {\left( {{\Gamma _m}} \right)^T}C$. The anti-symmetric inner product between two arbitrary spinors is defined as $\left( {\psi \chi } \right) \equiv \sum {{\psi ^\alpha }{C_{\alpha \beta }}{\chi ^\beta }} $, denoted by a parenthesis $()$. Our spinors satisfy a symplectic Majorana condition:
\begin{equation}\label{eq:symplectic_Majorana_condition}
  \overline {\xi _I}  = C \epsilon^{IJ} \xi _J.
\end{equation}
Here, $C$ is the antisymmetric charge conjugation matrix and $\overline{\xi_I}$ denotes complex conjugation. $SU(2)_R$ indices are raised and lowered with the invariant matrices $\epsilon^{IJ} = \left( \begin{smallmatrix} 0 & 1 \\ -1 & 0 \end{smallmatrix} \right)$ and $\epsilon_{IJ} = \left( \begin{smallmatrix} 0 & -1 \\ 1 & 0 \end{smallmatrix} \right)$ which act from the left.

  The following Fierz-identities are used
  \begin{equation}
    2{\xi _1}\left( {{\xi _2}{\xi _3}} \right) - 2{\xi _2}\left( {{\xi _1}{\xi _3}} \right) = {\xi _3}\left( {{\xi _2}{\xi _1}} \right) + {\Gamma ^m}{\xi _3}\left( {{\xi _2}{\Gamma _m}{\xi _1}} \right),
  \end{equation}
  \begin{equation}
    {\Gamma ^{mn}}{\xi _1}\left( {{\xi _2}{\Gamma _{mn}}{\xi _3}} \right) =  - 4\left[ {{\xi _3}\left( {{\xi _2}{\xi _1}} \right) + {\xi _2}\left( {{\xi _3}{\xi _1}} \right)} \right].
  \end{equation}

\paragraph{Hodge duality and the horizontal/vertical decomposition:}

A generic $p$-form can be decomposed into its horizontal and vertical components via
\begin{equation}\label{eq:horizontal-vertical_decomposition}
  \omega  = {\omega _H} + {\omega _V} \equiv {\iota _R}\left( {\kappa  \wedge \omega } \right) + \kappa  \wedge {\iota _R}\omega ,
\end{equation}
Horizontal two-forms $\omega_H$ can be further projected onto their self-dual and anti self-dual parts,
\begin{equation}\label{eq:self_duality_decomposition}
  \omega  = {\omega ^ + } + {\omega ^ - } \equiv \frac{1}{2}\left( {\omega  + {\iota _R}*\omega } \right) + \frac{1}{2}\left( {\omega  - {\iota _R}*\omega } \right),
\end{equation}
Using ${\left( {{\iota _R}*} \right)^2} = {R^m}{\kappa _m} = 1$ one sees that ${\iota _R}*{\omega ^ \pm } =  \pm {\omega ^ \pm }$.

Given a pair $\omega^\pm$ of self-dual and anti-self-dual 2-forms, one finds
  \begin{equation}
    \Omega_{mn} \equiv {\left( {{\omega ^ + }} \right)_{mk}}{\left( {{\omega ^ - }} \right)^k}_n - {\left( {{\omega ^ + }} \right)_{nk}}{\left( {{\omega ^ - }} \right)^k}_m = 0.
    \label{vanishing-product}
  \end{equation}
This can be easily shown in an orthonormal basis (and properly oriented): ${\left( {{\omega ^ \pm }} \right)_{12}} =  \pm {\left( {{\omega ^ \pm }} \right)_{34}}$, ${\left( {{\omega ^ \pm }} \right)_{13}} =  \pm {\left( {{\omega ^ \pm }} \right)_{42}}$, ${\left( {{\omega ^ \pm }} \right)_{14}} =  \pm {\left( {{\omega ^ \pm }} \right)_{23}}$, and for instance,
  \begin{equation}
    \begin{gathered}
      {\Omega _{13}} = \omega _{12}^ + \omega _{23}^ -  + \omega _{14}^ + \omega _{43}^ +  - \omega _{32}^ + \omega _{21}^ -  - \omega _{34}^ + \omega _{41}^ -  \hfill \\[0.5em]
      \;\;\;\;\;\;\; =  - \omega _{23}^ + \omega _{34}^ +  - \omega _{21}^ + \omega _{14}^ -  + \omega _{43}^ + \omega _{32}^ -  + \omega _{41}^ + \omega _{12}^ -  =  + \omega _{23}^ + \omega _{34}^ -  + \omega _{21}^ + \omega _{14}^ -  - \omega _{43}^ + \omega _{32}^ -  - \omega _{41}^ + \omega _{12}^ +  \hfill \\ 
    \end{gathered}
  \end{equation}
where the first and the second equality in the second line come from applying on the first line the self-duality on $\omega^+$ and the anti-self-duality on $\omega^-$.

For explicit calculations, we use
\begin{equation}\label{eq:Dirac_matrices}
  \Gamma^1 = - \sigma_1 \otimes \mathbb{I}, \qquad
  \Gamma^2 = \sigma_2 \otimes \sigma_1, \qquad
  \Gamma^3 = \sigma_2 \otimes \sigma_2, \qquad
  \Gamma_4 = \sigma_2 \otimes \sigma_3, \qquad
  \Gamma_5 = \sigma^3 \otimes \mathbb{I}.
\end{equation}
The charge conjugation matrix is $C = \Gamma^{24}$.

\section{The spinorial holomorphy condition}\label{sec:appendix_spinorial_holomorphy_conditions}

  We prove the spinorial characterization of $T^{1,0}$ and $T^{1,0} \oplus \mathbb{R}R$ in equations \eqref{spinorial-(1,0)} and \eqref{spinorial-THF}. Assume $X \in T^{1,0}$, namely ${\Phi ^m}_n{X^n} = i{X^m}$. Then
  \begin{equation}
    - \sqrt {\frac{1}{{\det \mathfrak{m}}}} {\mathfrak{m}_I}^J\left( {{\xi ^I}{\Gamma ^m}{\Gamma _n}{\xi _J}} \right){X^n} + i\delta _I^J\left( {{\xi ^I}{\Gamma ^m}{\Gamma _n}{\xi _J}} \right){X^n} = 0,
  \end{equation}
which simplifies to ${H_I}^J\left( {{\xi ^I}{\Gamma ^m}{\Gamma _n}{\xi _J}} \right){X^n} = 0$, where ${H_I}^J \equiv {\left( {\det \mathfrak{m}} \right)^{ - 1/2}}{\mathfrak{m}_I}^J - i\delta _I^J$.

  Due to the reality properties of $\mathfrak{m}_{IJ}$ we have the identity $\sum {{H_I}^K\overline {{H_J}^K} }  = 2i{H_I}^J$. Contracting the above with $\overline {{X^m}} $ and inserting the identity, one has
  \begin{equation}
    \begin{gathered}
    \;\;\;\;\overline {{X_m}} {H_I}^K\overline {{H_J}^K} {\varepsilon ^{II'}}\xi _{I'}^\alpha {C_{\alpha \beta }}{\left( {{\Gamma ^m}} \right)^\beta }_\gamma {\left( {{\Gamma _n}} \right)^\gamma }_\delta \xi _J^\delta {X^n} = 0 \hfill \\[0.5em]
    \Leftrightarrow \overline {{X^m}{H_K}^I{{\left( {{\Gamma _m}} \right)}^\gamma }_\beta \xi _I^\beta } {X^n}{H_K}^J{\left( {{\Gamma _n}} \right)^\gamma }_\delta \xi _J^\delta  = 0 \hfill \\[0.5em]
    \Leftrightarrow \sum\limits_{K,\alpha } {\Delta _K^\alpha \overline {\Delta _K^\alpha } }  = 0 \hfill \\ 
    \end{gathered} 
  \end{equation}
This implies $\Delta _K^\alpha  = 0$, namely ${H_I}^J{X^m}{\Gamma _m}{\xi _J} = 0$.

It is obvious how to extend to $X \in T^{1,0} \oplus \bC R$: One just needs to project out the vertical components of $X$, and the horizontal components should satisfy ${H_I}^J{X^m}{\Gamma _m}{\xi _J} = 0$. Namely,
  \begin{equation}
    {H_I}^J{\Pi ^m}_n{X^n}{\Gamma _m}{\xi _J} = 0, \;\;\;\;{\Pi ^m}_n = \delta _m^n - {R^m}{\kappa _n}.
  \end{equation}

\section{The Nijenhuis tensor, $\pounds_{sR} \Phi$ and integrability of THFs}

In this appendix we discuss the integrability of the canonical almost CR structure $\Phi = \Phi[t]$ in terms of its Niejenhuis tensor.

\subsection{The Nijenhuis tensor and $[T^{1,0}, T^{1,0}]$}

Given an almost CR structure $\left( {\kappa ,R,\Phi } \right)$, one can define its Nijenhuis tensor as
\begin{equation}\label{eq:Niejenhuis_w-out_indices}
  {N_\Phi }\left( {X,Y} \right) \equiv  - \left[ {X,Y} \right] + \kappa \left( {\left[ {X,Y} \right]} \right)R + \left[ {\Phi X,\Phi Y} \right] - \Phi \left[ {\Phi X,Y} \right] - \Phi \left[ {X,\Phi Y} \right],
\end{equation}
which can be expressed in components
\begin{equation}
  {N^k}_{mn} \equiv {\Phi ^l}_m{\nabla _l}{\Phi ^k}_n - {\Phi ^l}_n{\nabla _l}{\Phi ^k}_m + {\Phi ^k}_l{\nabla _n}{\Phi ^l}_m - {\Phi ^k}_l{\nabla _m}{\Phi ^l}_n.
  \label{Nijenhuis}
\end{equation}
For simplicity, we restrict our analysis to the canonical almost CR structure determined by $t_{IJ}$, namely
  \begin{equation}
    {\Phi ^m}_n \equiv \sqrt {\frac{1}{{\det t}}} {t^{IJ}}\left( {{\xi _I}{\Gamma ^m}_n{\xi _J}} \right).
  \end{equation}
By explicitly inserting the Killing spinor equation and the dilatino equation into \eqref{Nijenhuis}, one finds that
\begin{equation}\label{eq:Niejenhuis_integrability}
  {N_\Phi }\left( {X,Y} \right) + d\kappa \left( {\Phi X,\Phi Y} \right)R = 0, \qquad
  \forall X,Y \in \Gamma(TM_H),
\end{equation}
provided that
\begin{equation}
  {X^m}{D_m} \left(\frac{t_{IJ}}{\sqrt{\det t}}\right) = 0, \qquad \forall X \in \Gamma(TM_H),
\end{equation}
where $TM_H$ is the horizontal part of the tangent bundle. Of course this condition is the same as in \eqref{eq:THF_integrability_obstructions}.

We will now show that the above condition \eqref{eq:Niejenhuis_integrability} is equivalent to the statement that
\begin{equation}
  \left[ {{T^{1,0}},{T^{1,0}}} \right] \subset {T^{1,0}} \oplus \bC R.
\end{equation}
To do so, consider $X, Y \in T^{1,0}$. Using $\Phi(X) = \imath X$ and $\kappa([X, Y]) = -d\kappa(X, Y)$, one can evaluate \eqref{eq:Niejenhuis_w-out_indices}:
\begin{equation}
  N_\Phi (X, Y) + d\kappa (X, Y) = -2 (1 + \imath \Phi) [X, Y] = -2 [X, Y]^{0,1}.
\end{equation}
It is clear that \eqref{eq:Niejenhuis_integrability} implies that $[X, Y] \in T^{1,0} \oplus \bC R$ and vice versa.

\subsection{$\pounds_{sR} \Phi$ and $[T^{1,0}, R]$}

In section \ref{sec:implications_for_localization} we showed gravitino and dilatino equations imply that for the canonical almost CR structure $\pounds_{sR} \Phi = 0$. For any $X \in TM$ it follows that
\begin{equation}
  \pounds_{sR} (\Phi X) = \Phi (s [R, X] - X(s) R) = s \Phi ([R, X]).
\end{equation}
On the other hand
\begin{equation}
  \pounds_{sR} (\Phi X) = [sR, \Phi X] = s [R, \Phi X] - (\Phi X)(s) R
\end{equation}
and thus
\begin{equation}
  s \Phi ([R, X]) = s [R, \Phi X] - (\Phi X)(s) R,
\end{equation}
which we rewrite as
\begin{equation}
  [R, \Phi X] = \Phi([R, X]) + (\Phi X) (\log s) R.
\end{equation}

Now, consider that any $X^{1,0} \in T^{1,0}$ can be written as $X^{1,0} = X - \imath \Phi X$ for some $X \in TM_H$. Then
\begin{equation}
  [X^{1,0}, R] = (1 - \imath \Phi)[X, R] + \imath (\Phi X) (\log s) R = [X, R]^{1,0} + (\kappa([X, R]) + \imath (\Phi X) (\log s)) R \in T^{1,0} \oplus \bC R.
\end{equation}
In other words, we have confirmed that the canonical almost CR structure defines a THF as long as the triplet $t_I^{\phantom{I}J}$ is covariantly constant; i.e.~equation \eqref{eq:THF_integrability_obstructions}.

\bibliographystyle{ytphys}
\bibliography{reference}

\end{document}